\documentclass[%
 aip,
 sd,%
 amsmath,amssymb,
 reprint,%
]{revtex4-1}

\usepackage{graphicx}
\usepackage{dcolumn}
\usepackage{bm}
\usepackage{CJK}
\usepackage{amsthm}
\begin{document}

\begin{CJK*}{GB}{}
\title{A practical theoretical formalism for atomic multielectron processes: direct multiple ionization
by a single Auger decay or by impact of a single electron or photon}%
\author{Pengfei Liu$^1$, Jiaolong Zeng$^{1,3,\bf \dagger}$\footnotetext{${\bf \dagger}$
To whom correspondence should be addressed. E-mail: jlzeng@nudt.edu.cn}, and Jianmin Yuan}
\email{jlzeng@nudt.edu.cn}
\address{College of Science, National University of Defense Technology, Changsha Hunan 410073, P. R. China}
\address{Graduate school of China Academy of engineering Physics, Beijing 100193, P. R. China}
\address{IFSA Collaborative Innovation Center, Shanghai Jiao Tong University, Shanghai 200240, P. R. China}
\date{\today}

\begin{abstract}
Multiple electron processes occur widely in atoms, molecules,
clusters, and condensed matters when they are interacting with
energetic particles or intense laser fields. Direct multielectron
processes are the most involved among the general multiple electron processes
and are the most difficult to describe theoretically. In this work, a unified
and accurate theoretical formalism is proposed on the direct multielectron processes
of atoms including the multiple Auger decay and multiple ionization by an impact of an
energetic electron or a photon based on the atomic collision theory described by a
correlated many-body Green's function. Such a practical treatment is made possible due to different coherence
features of the particles (matter waves) in the initial and final states. We first explain how
the coherence characteristics of the ejected continuum electrons is
largely destructed, by taking the electron impact direct double
ionization process as an example. This process is completely different from the single
ionization where the complete interference can be maintained. The detailed expressions are
obtained for the energy correlations among the continuum electrons and energy
resolved differential and integral cross sections according to the
separation of knock-out and shake-off mechanisms for the electron
impact direct double ionization, direct double and triple Auger
decay, and double and triple photoionization processes. Extension to
higher-order direct multielectron processes than triple ionization
is straight forward by adding contributions of following
knock-out and shake-off processes. The approach is applied to
investigate the electron impact double ionization processes of
C$^+$, N$^+$, and O$^+$, the direct double and triple Auger decay of
the K-shell excited states of C$^{+}$ $1s2s^22p^2$ $^2D$ and $^2P$,
and the double and triple photoionization of lithium. Comparisons
with available experimental and theoretical results show that our proposed
theoretical formalism is accurate and effective in treating the
atomic multielectron processes.
\end{abstract}
\maketitle
\end{CJK*}
\section{Introduction}

Multielectron processes occuring in a single Auger decay and in collision of
atoms (including atomic ions), molecules, clusters, and condensed matters with a single photon or a single charged particle
belong to one of the most interesting areas of physics. They are
vital to understand the nature of many-electron and
many-particle transitions and play a significant role in a variety of
practical applications such as modeling of plasma processes,
astrophysics and charge state distribution and evolution of atoms
exposed to an electron beam \cite{Shevelko} or a radiation field.
Multielectron processes can be classified as direct and indirect
ones, where multiple electrons are emitted simultaneously or sequentially.
The indirect processes are determined by removal of
an inner-shell electron and subsequent autoionization, where only
single electron transitions are involved in each step. Direct
processes, however, require sophisticated theoretical methods which
should come out of the framework of one-electron approximation.

The correlated motion of electrons in the direct multielectron
processes (DMEP) has been a concern of physics since the early days
of quantum mechanics \cite{Tanner}. Double photoionization (DPI)
\cite{nat2004-2,sci2007,Knapp,Schoffler,AHLiu}, electron impact
double ionization (EIDI)
\cite{Lahmam,Ford,Taouil,Marji,Dorn,Boeyen,Durr} and double Auger
decay (DAD) \cite{Carlson,Viefhaus,Lablanquie,Hikosaka} are the
lowest order of such processes, where two electrons are
simultaneously ejected into the continuum state by
impact of a single photon or a single electron or by a single Auger decay. A wealth of information on
the physical mechanisms, angular distributions and energy
correlations between the ejected electrons has provided deeper
insight into these electron-electron interaction dynamical
processes. Understanding the correlation can help to control the
multielectron processes \cite{Hogle}. The latest attractive research
is the direct triple processes as demonstrated theoretically by
Colgan {\it et al.} \cite{Colgan2,Colgan3} for triple
photoionization and by M\"{u}ller {\it et al.} \cite{Muller,Muller6} for triple Auger decay, the authors
of the latter work experimentally observed unambiguously a four-electron
Auger decay with simultaneous emission of three electrons.

Theoretical description on these correlated multielectron processes,
however, poses a formidable challenge to the theorists, even for
the simplest test cases of the three-body Coulomb problems of small
systems such as the helium atom or the hydrogen molecules
\cite{Tanner,Lambropoulos}. Nearly exact calculations for such
systems are available only for bound states, whereas the correlated
motion of electrons in DMEP is still not fully understood. A number of
sophisticated calculations have been performed for DPI
\cite{Vanroose,Hart,Colgan,Istomin,Ivanov,Singh}, EIDI
\cite{Bray,Kheifets,Dorn2,Pindzola,Mengoue,Li} and DAD
\cite{Amusia2,Pindzola5,Pindzola6}. These theoretical calculations
investigated the many-fold differential or integral cross sections by
numerically solving the time-dependent Schr\"{o}dinger equation or
by treating the electron-electron interaction as a perturbation to
investigate the double continuum process. Most of the calculations deal
with the simpler systems such as helium atom and hydrogen molecule, and
only a few on Li- and Be-like atoms or atomic ions. For more complex
multi-electron atoms, however, such sophisticated calculations are
difficult or untractable even for the double continuum process, let
alone for the triple and higher order multielectron processes.

As is well known, the electron correlation is a prerequisite for a
DMEP to occur, i.e., the highly correlated
initial state wave-function and correlated motion in the final state
with the exchange of energy between the outgoing electrons. However,
the features of electron correlation in the final state are
different from that of the initial state in that the first ejected
continuum electron is coupled with the remaining environment by the
Coulomb interaction, which results in the loss of partial
information of the wave phases \cite{sci2007}. By fully taking advantage of the different correlation features, the theoretical
description can be greatly simplified with little loss of
computation accuracy. The central physical thought is that the interference between
different channels (see Fig. 1 for such a pathway, taking a direct quadruple ionization as an example) is largely destructed due to
the continuous variations of wave phases in the transition matrix elements.
The more ejected electrons, the less the interference effects remain.

In this work, we proposed a practical and accurate theoretical formalism on the
DMEPs including multiple Auger decay and multiple ionization by impact of a single photon
or electron. Detailed expressions have been obtained for the energy resolved differential
and integral cross sections for the direct double and triple processes. The calculations
of the fully (angular) differential cross section can also be implemented according to
the separation of ionization mechanism.
Theoretical treatment for the higher order DMEP than triple processes is even more
challenging. To the best of our knowledge, no report on such processes is available
in the literature. However, extension of our theoretical formalism to such higher order
multiple ionization processes is straight forward. Moreover, most past theoretical investigations on multiple
ionization have dealt with simple atomic systems of few electrons such as helium and lithium.
Yet our solution can treat any complex heavy Z atoms with many electrons.
\begin{figure}[htbp]
\begin{center}
\includegraphics[width=\columnwidth]{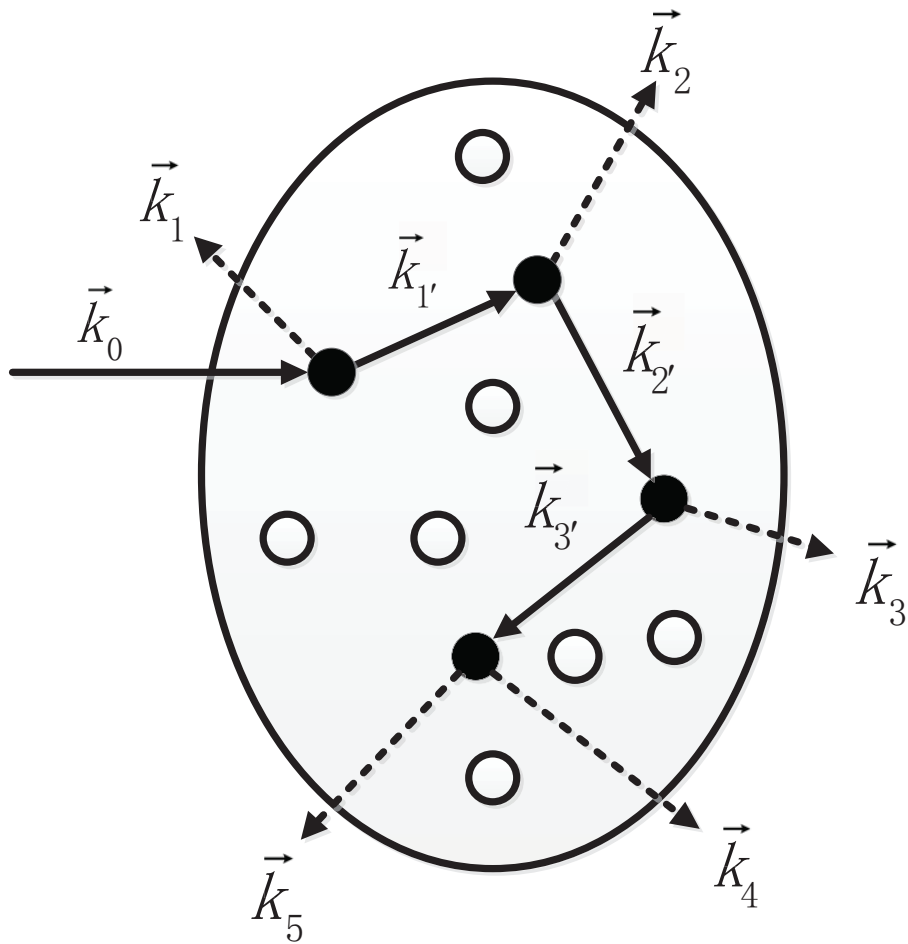}
\caption{A schematic illustration of a channel in the atomic direct multielectron processes,
taking a quadruple ionization as an example. The momentum of incoming projectile is denoted as
$\vec{k}_0$ and each circle represents a possible electronic state. Along with the pathway, all solid
circles constitute a reaction channel. Any solid circles can be replaced by open circles to form another channel.}
\end{center}
\end{figure}
\section{Theoretical formalism}
\begin{figure}[htbp]
\begin{center}
\includegraphics[width=\columnwidth]{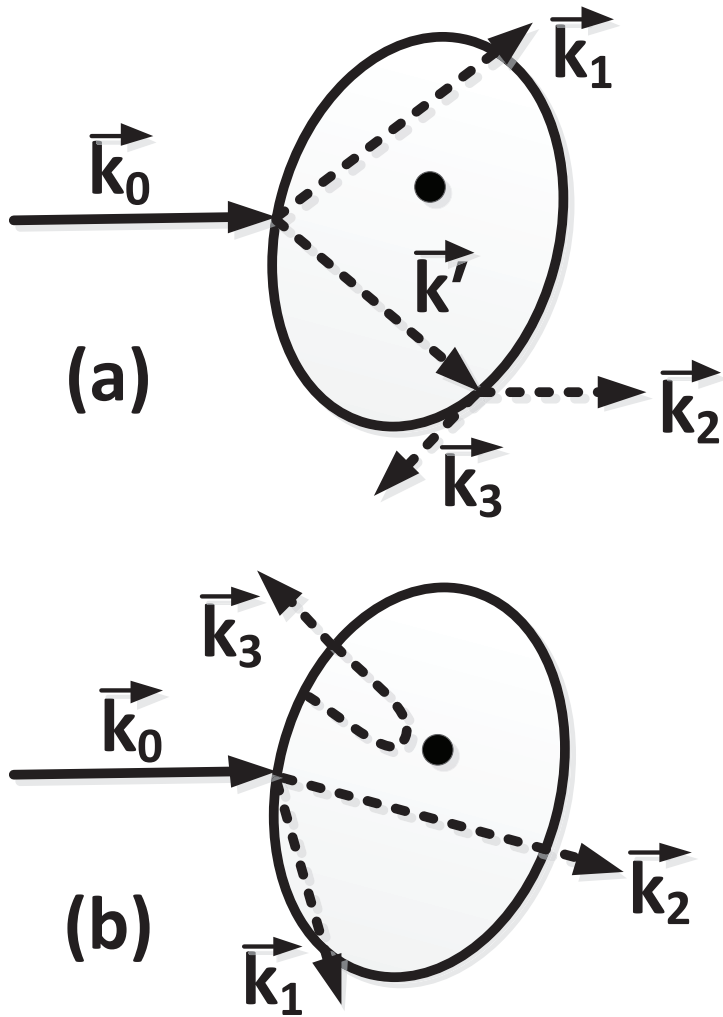}
\caption{A schematic representation of the mechanisms of (a) knock-out (KO) and (b) shake-off (SO)
in electron impact direct double ionization processes.
The momentum vector of the incoming projectile is denoted by an arrow labeled by $\vec{k}_0$,
and the momenta of the three ejected electrons are denoted by $\vec{k}_1$, $\vec{k}_2$, and $\vec{k}_3$.
The nuclei of the atom is shown as a full dot.}
\end{center}
\end{figure}
First we take the electron impact direct double ionization [also
called (e,3e)] processes of an isolated many-electron atom as an
example to start our discussion. The complete dynamics, which is described by
eight-fold differential cross section (DCS), is spanned over the
momentum space of the three continuum electrons where eight
independent variables are required considering the energy
conservation relation \cite{Berakdar}
\begin{equation}
\frac {d\sigma_{if}(E_0,E_1,E_2,\Omega_1,\Omega_2,\Omega_3)}{dE_1dE_2d\Omega_1d\Omega_2d\Omega_3}=
(2\pi)^4 \mu^2 \frac {k_1k_2k_3}{k_0} |T_{if}|^2
\end{equation}
where $i$ and $f$ represent the initial and final states, $k_0$ and
$E_0$ the momentum and kinetic energy of the incident electron,
$E_p$, $k_p$, and $\Omega_p$ ($p$=1, 2, 3) the kinetic energy,
magnitude of momentum, and solid angle of the three outgoing
electrons, and $\mu$ the reduced mass of the incident electron and
the target. The transition probability amplitude $T_{if}$, which defines a Lippmann-Schwinger equation for
the transition matrix element, can be expressed in terms of the Green's function
\begin{equation}
T_{if} = <\vec{k}_1,\vec{k}_2,\vec{k}_3,\varphi_f|V_f(1+G^+V_i)|\vec{k}_0,\psi_i>
\end{equation}
where $|\vec{k}_0,\psi_i>$ ($V_i$) and
$|\vec{k}_1,\vec{k}_2,\vec{k}_3,\varphi_f>$ ($V_f$) are the initial
and final state vectors (transition potential operators), $G^+$ the
fully correlated four-body Green's function, which is the resolvent
of the whole physical system of incident electron plus target with
appropriate boundary conditions. The state vectors satisfy the
time-independent Schr\"{o}dinger (or relativistic quantum mechanics) equation
$H_i|\vec{k}_0,\psi_i>=E_i|\vec{k}_0,\psi_i>$ and
$H_f|\vec{k}_1,\vec{k}_2,\vec{k}_3,\varphi_f>=E_f|\vec{k}_1,\vec{k}_2,\vec{k}_3,\varphi_f>$
with $H_i$ and $H_f$ being the Hamiltonians of the initial
projectile-atom system and of the three charged continuum electrons
in the field of a doubly charged ion and $E_i$ and $E_f$ the
energies of the initial and final state. In the description of final state, the designation
of electron 1, 2, and 3 is introduced for convenience, yet there is no other special meaning as these electrons are ejected simultaneously
in the direct ionization processes. In both the initial and final wave functions,
the effects of all potentials including the exchange interaction potential are
included in the calculations and hence they should be treated in the same footing.
The exchange potential operators $V_i$
and $V_f$ describe the Coulomb interaction between the incident
electron and the target before collision and between the three outgoing electrons and
the final state $\varphi_f$ after collision. Assuming non-interacting asymptotic
states after collision $V_f$ can be expressed as
\begin{equation}
V_f=V_{e_1e_2}+V_{e_1e_3}+V_{e_2e_3}+V_{e_1f}+V_{e_2f}+V_{e_3f}
\end{equation}
\begin{figure}[t]
\begin{center}
\includegraphics[width=\columnwidth]{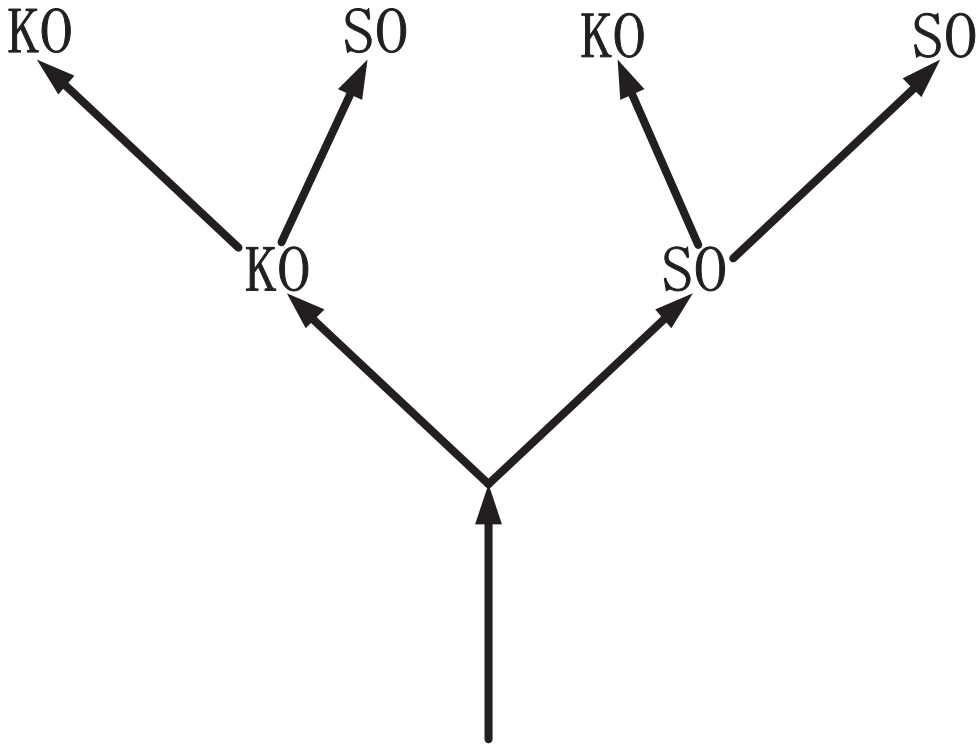}
\caption{ A schematic diagram of the physical mechanisms following the KO and SO processes of the DDAD
in the DTAD. It is straight forward to extend it to higher order multielectron processes.}
\end{center}
\end{figure}

Considering the intermediate states after single ionization [see Fig. 2(a)],
all bound states and continuum states of the projectile-target system form a
complete set and thus satisfy the closure condition
\begin{equation}
\sum\limits_{n} {\int\limits_{\vec{k'}}}\llap{$\sum$} d\vec{k'}
|\vec{k}_1,\vec{k'},\chi_n><\vec{k}_1,\vec{k'},\chi_n|=1
\end{equation}
where $|\chi_n>$ is a (quasi-)bound state of the intermediate ion
with one higher charge. In the above equation, we number one of the continuum electron
as "1" for the convenience of expression. As we pointed out in the above that the simultaneously
ejected electrons are indistinguishable and thus the designation of electron "1" has no special
role. The summations over the intermediate middle states include a summation
over all possible bound states of the target and a summation and integration over a complete
set of bound and continuum states of one more highly ionized target.
Inserting this identity into the probability amplitude $T_{if}$ one
obtains
\begin{equation}
T_{if} = \sum\limits_{n} {\int\limits_{\vec{k'}}}\llap{$\sum$} <\vec{k}_1,\vec{k}_2,\vec{k}_3,\varphi_f|V_f
|\vec{k}_1,\vec{k'},\chi_n> d\vec{k'}<\vec{k}_1,\vec{k'},\chi_n|(1+G^+V_i)|\vec{k}_0,\psi_i>
\end{equation}

Let us denote
$<\vec{k}_1,\vec{k}_2,\vec{k}_3,\varphi_f|V_f|\vec{k}_1,\vec{k'},\chi_n>=a(\vec{k'})e^{i\phi_a(\vec{k'})}$
and
$<\vec{k}_1,\vec{k'},\chi_n|(1+G^+V_i)|\vec{k}_0,\psi_i>=b(\vec{k'})e^{i\phi_b(\vec{k'})}$,
where $a(\vec{k'})$ and $b(\vec{k'})$ represent the absolute
magnitudes of the respective matrix element and $\phi_a(\vec{k'})$
and $\phi_b(\vec{k'})$ denote their phases. For a particular final state of
$|\vec{k}_1,\vec{k}_2,\vec{k}_3,\varphi_f>$, $\phi_a(\vec{k'})$ and
$\phi_b(\vec{k'})$ vary with $\vec{k'}$ continuously from the lowest
to the highest possible values and therefore the interference between
different channels denoted by $\vec{k'}$ is largely destructed by
the integrations over $\vec{k'}$ when we calculate the cross section
between the initial and final states. The possible nonzero interference
would very highly sensitively depend on some particular values of $\vec{k}_1$, $\vec{k}_2$,
and $\vec{k}_3$, and the integrations over the directions of
$\vec{k}_2$ and $\vec{k}_3$ will average out the existing
interference effects when we calculate the energy correlated
differential cross sections or the total cross sections. The last
possibility of the missing of interference effects is the insufficient high
enough angular resolution of the momenta $\vec{k}_2$ and
$\vec{k}_3$ measurement even for a complete momentum differential
cross section. As we know, the high sensitivity on particular values of
momenta of three continuum electrons requires a very high resolution in the experiment,
which is perhaps beyond the capability of our most advanced instrument
or is limited by the physical principle.

It is also helpful to express the probability amplitude [Eq. (2)] as
a multiple scattering series, which can be derived by iterating the
integral Lippmann-Schwinger equation of the Green operators
\cite{Berakdar}. If we consider only the second order terms, the
dominant mechanisms of shake-off (SO) and knock-out (KO) [which is
sometimes called two step 2 and two step 1 in the literature] can be
identified. KO describes the correlated dynamics of the three
continuum electrons, where the incident electron ionizes a secondary
electron and then either one knocks out the third electron in an
inelastic scattering process. A removal of the second
electron causes a sudden change of atomic potential and a third electron
can be shaken off to the continuum state, which is called SO mechanism. In the multiple scattering
series, there are many terms which can be categorized into the
mechanisms of either KO [see Fig. 2(a)] or SO [see Fig. 2(b)].
According to the separation of KO and SO mechanisms, the angular differential
cross sections can be obtained. After
integrating the eight-fold DCS over the solid angles of all
electrons, two-fold DCS based on the two mechanisms reads
\begin{equation}
\frac {d\sigma_{if}^{KO}(E_0,E_1,E_2)}{dE_1dE_2}=\frac {1}{\pi} \sum\limits_{n}\frac {d\sigma_{in}(E_0,E_1)}{dE_1} \frac {d\sigma_{nf}(E_1,E_2)}{dE_2}
\end{equation}
\begin{equation}
\frac {d\sigma_{if}^{SO}(E_0,E_1,E_2)}{dE_1dE_2}=\frac {1}{\pi} \sum\limits_{n}\frac {d\sigma_{in}(E_0,E_1)}{dE_1} |<E_2,E_1-I_n-E_2,\varphi_f|E_1,\chi_n>|^2
\end{equation}
where $I_n$ is the ionization energy of state $|\chi_n>$, $\frac
{d\sigma_{in}(E_0,E_1)}{dE_1}$ and $\frac
{d\sigma_{nf}(E_1,E_2)}{dE_2}$ are, respectively, the
electron-impact single ionization DCS from $|\psi_i>$ to
intermediate state $|\chi_n>$ and from $|\chi_n>$ to the final state
$|\varphi_f>$. The integral cross section can be obtained by
integrating over the kinetic energies of continuum electrons
\begin{equation}
\sigma_{if}^{KO}(E_0)=\frac {1}{2\pi} \sum\limits_{n}\sigma_{in}(E_0) \int\limits_0^{E_0-I_i} f(E_0,E_1) \sigma_{nf}(E_1)dE_1
\end{equation}
\begin{equation}
\sigma_{if}^{SO}(E_0)=\frac {1}{2\pi} \sum\limits_{n}\sigma_{in}(E_0) \int\limits_0^{E_0-I_i} dE_1 f(E_0,E_1)\int\limits_0^{E_0-I_i-I_n}|<E_2,E_1-I_n-E_2,\varphi_f|E_1,\chi_n>|^2 dE_2
\end{equation}
where $I_i$ and $I_n$ being the
ionization potentials of states $|\psi_i>$ and $|\chi_n>$. A factor
of $\frac {1}{2}$ is introduced to avoid double counting of the
contribution. The function $f(E_0,E_1)$ describes the population
fraction per unit energy interval of $E_1$ in the second collision,
which is defined as $\frac{d\sigma_{in}(E_0,E_1)}{dE_1}=\sigma_{in}(E_0)f(E_0,E_1)$.

The above formulas of KO and SO mechanisms are readily applicable to
the DPI and DAD processes by letting $f(E_0,E_1)=1$ and replacing
the electron impact ionization cross section $\sigma_{in}(E_0)$ (the
first term in the right hand) by the single photoionization cross
section or single Auger decay rate, respectively. For example, the
DAD rates due to KO and SO mechanisms reads as
\begin{equation}
A^2_{if}(KO)=\frac {1}{\pi} \sum\limits_{n}A^1_{in} \sigma_{nf}(E_0)
\end{equation}
\begin{equation}
A^2_{if}(SO)=\frac {1}{\pi} \sum\limits_{n}A^1_{in} \int\limits_0^{E_0-I_i}|<E_1,E_0-I_n-E_1,\varphi_f|E_0,\chi_n>|^2 dE_1
\end{equation}
where $A^1_{in}$ is the single Auger decay rate from the initial
state $|\psi_i>$ to a middle state $|\chi_n>$. Note, however, that
for the DPI process the electron impact cross section in the second step of
KO should include only those contributions which fulfill the dipole
selection rule. Past work showed that the separate formulation of KO
and SO mechanisms offers an accurate description for the DPI
\cite{Hoszowska,Huotari,Knapp,Schneider,Pattard} and DAD
\cite{Amusia2,Zeng,Zeng2,Liu}. The separability of the two-step
expressions in the KO and SO indicates that the ejected continuum
electrons are not as highly correlated as the bound electrons in
the initial state. The Coulomb interaction with the second electron
and remaining atomic ion modifies the matter wave phases of the
first electron and thus induces the loss of partial phase
information. This is a common phenomenon for the direct
multi-electron processes and hence a unified theoretical
description is feasible. The unified theory bridges these
direct multiple processes and reveals such connections. It also
offers a deeper insight into the similarity between these physical
processes, which has been demonstrated between the DPI and electron
impact single ionization processes \cite{Taouil,Schneider}. Further
experimental and theoretical work should be able to demonstrate the
similarity between the Auger decay and ionization processes by a single photon or a single electron.

The above method can readily be extended to the higher order
multielectron processes. Take the direct triple Auger decay (DTAD) process
as an example to illustrate the procedure. According to the
separation of physical mechanisms, DTAD can proceed via the two
routes of KO and SO which results in four pathways (see Fig. 3).
Following the KO mechanism of DAD, there are also two pathways of KO
and SO with the decay rate being expressed as

\begin{widetext}
\begin{equation}
A^3_{if}(KO+KO)=\frac {1}{2\pi^2}\sum\limits_{j}\sum\limits_{n}A_{ij}^1 \sigma_{jn}(E _0)
 \int\limits_0^{E_0-I_j} f(E_0,E_1) \sigma_{nf}(E_1)dE_1
\end{equation}
\begin{equation}
A^3_{if}(KO+SO)=\frac {1}{2\pi^2}\sum\limits_{j}\sum\limits_{n}A_{ij}^1 \sigma_{jn}(E _0)\int\limits_0^{E_0-I_j}dE_1
f(E_0,E_1)  \int\limits_0^{E_0-I_j-I_n}|<E_2,E_1-I_n
-E_2,\varphi_f|E_1,\chi_n>|^2 dE_2
\end{equation}
\end{widetext}
Comparing with the KO mechanism in the DAD, we term these two processes as a double KO
(KO+KO) and a KO and SO (KO+SO). The corresponding expressions
following the SO can be similarly obtained, which results in the
mechanisms of what are termed as a double SO (SO+SO) and SO+KO
\begin{widetext}
\begin{equation}
A^3_{if}(SO+KO)=\frac {1}{2\pi^2}\sum\limits_{j}\sum\limits_{n}A_{ij}^1 \int\limits_0^{E_0-I_j}dE_1| <E_1,E_0-I_j-E_1,\varphi_n|E_0,\psi_j>|^2 \sigma_{nf}(E_1)
\end{equation}
%
%
\begin{equation}
A^3_{if}(SO+SO)=\frac {1}{2\pi^2}\sum\limits_{j}\sum\limits_{n}A_{ij}^1 \int\limits_0^{E_0-I_j}dE_1
|<E_1,E_0-I_j-E_1,\varphi_n|E_0,\psi_j>|^2
\int\limits_0^{E_0-I_j-I_n}|<E_2,E_1-I_n-E_2,\varphi_f
|E_1,\chi_n>|^2 dE_2
\end{equation}
\end{widetext}
The above four formulas of DTAD are also applicable for the direct triple photoionization by replacing the
single Auger decay rate $A_{ij}^1$ (the first term in the right hand) by the single
photoionization cross section.

The required quantities including the single Auger decay rate,
electron- and photon-impact single ionization cross section are
obtained using the Flexible Atomic Code developed by Gu \cite{Gu2}
and the R-matrix method \cite{Berrington,Liuyp,Liuyp2,Liuyp3}. The
theory on these single electron processes can be found elsewhere
\cite{Zeng,Zeng2,Liu,Liuyp,Liuyp2,Liuyp3,Zeng11,LiuPF11,LiuPF12} and
therefore the details are not given here. To obtain as accurate
results as possible, electron correlations should be adequately
considered in each step \cite{Zeng3,Zeng4}.

Here we do not give the expressions of higher order multi-electron
processes such as direct tetrad Auger decay, electron impact direct
triple ionization and tetrad photoionization, yet extension of the
approach to these processes are straight forward. It should also be
extendable to more complex physical systems such as molecules and
clusters. The present theoretical formalism is helpful to understand other
particle-particle many-body scattering problem \cite{Abdurakhmanov}
and long-range interactions between ultracold atoms and molecules
\cite{Lepers}.
\section{Results and discussions}\label{sec:endnotes}
In what follows, we apply the unified theoretical formalism to
investigate three typical multielectron processes of multiple Auger
decay and multiple ionization by impact of a single photon or
a single electron with atoms or atomic ions. First we deal with the electron
impact double ionization processes.
\subsection{Electron impact double ionization}
Atomic helium and helium-like ions are the simplest systems to
investigate the electron impact double ionization processes
\cite{Durr,Ren,Gasaneo}. For such simple systems, only direct processes are
possible. Time-dependent close-coupling (TDCC) method has been applied
to these systems \cite{Pindzola21,Pindzola22,Pindzola23,Pindzola24}, which have provided a
good description for the double ionization cross sections. Here we
deal with more complex systems of C$^+$, N$^+$, O$^+$,
which still lack accurate theoretical explanations for the measured
experimental results \cite{Lecointre,Zambra,Westermann}. For these ions, no
close-coupling calculations have been carried out up to now.
Recently, Jonauskas {\it et al.} \cite{Jonauskas2014} investigated
the electron impact double ionization cross section of C$^+$ and O$^+$
by presenting the process as a sequence of two- and three-step
processes arising from ionization-ionization,
ionization-excitation-ionization, and
excitation-ionization-ionization processes.

\begin{figure}[htbp]
\begin{center}
\includegraphics[width=\columnwidth]{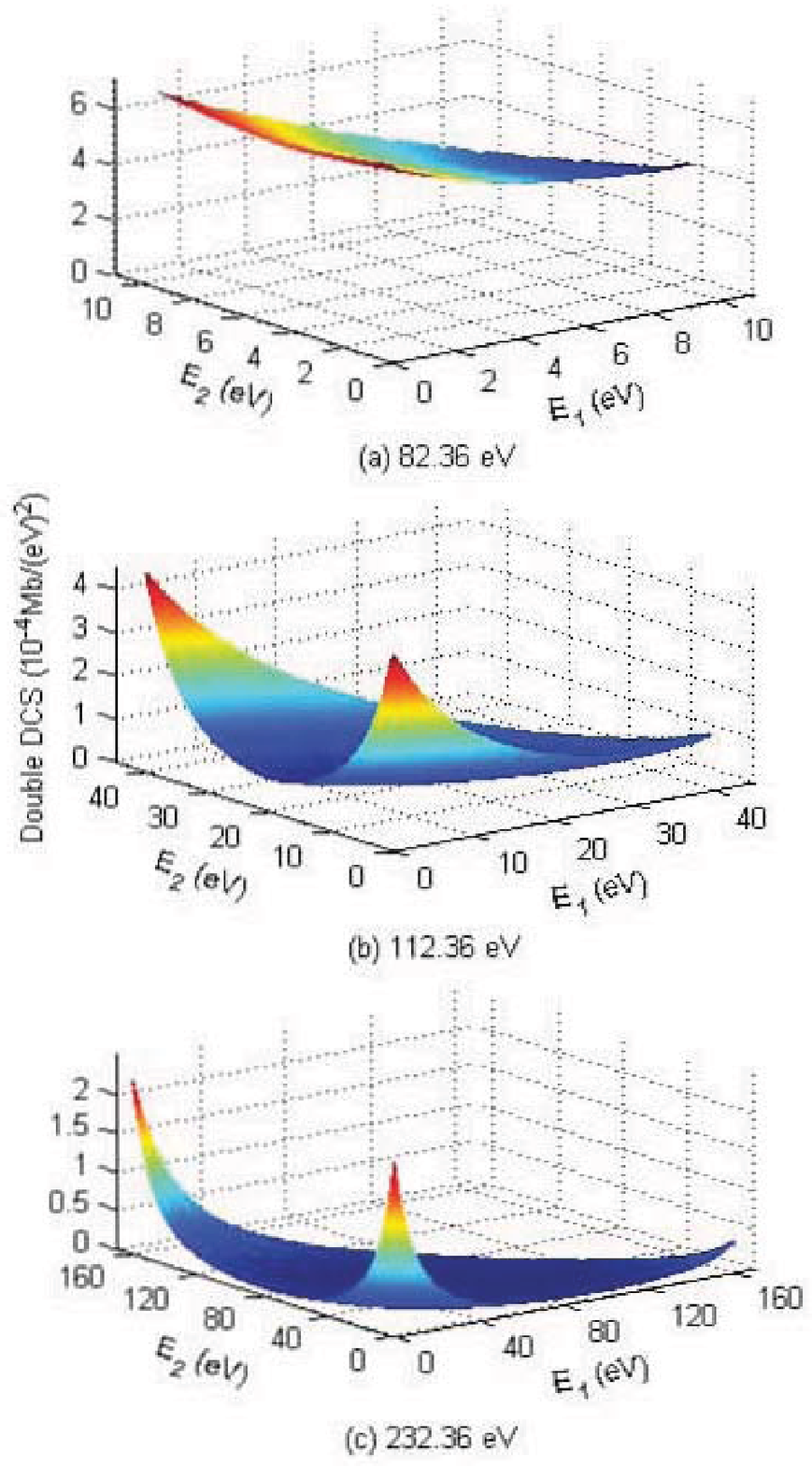}
\caption{ Energy resolved double differential cross section (DCS) of the ground term of C$^+$ at incident electron energies of
(a) 82.36 eV,  (b) 112.36 eV and (c) 232.36 eV, which are above the
double ionization threshold by 10.0, 40.0, and 160.0 eV, respectively, for the pathway of
e+C$^{+}$ $1s^22s^22p$ $^2P^o$ $\rightarrow$ C$^{2+}$ $1s^22s^2$ $^1S$+2e $\rightarrow$ C$^{3+}$ $1s^22s$ $^2S$ +3e
in the electron impact direct double ionization process.}
\end{center}
\end{figure}

First we investigate the energy resolved double DCS which reflects
the energy correlations in the electron impact direct double
ionization. Figure 4 shows the double DCS for a pathway of
e+C$^{+}$ $1s^22s^22p$ $^2P^o$ $\rightarrow$ C$^{2+}$ $1s^22s^2$
$^1S$+2e $\rightarrow$ C$^{3+}$ $1s^22s$ $^2S$ +3e at the incident
electron energy of 82.36, 112.36 and 232.36 eV, which are above the
double ionization threshold by 10.0, 40.0, and 160.0 eV, respectively. The
single and double ionization potentials of C$^{+}$ were calculated to be
24.96 and 72.36 eV, which is in good agreement with the experimental
values \cite{NIST} of 24.38 and 72.27 eV (see table 1). The
theoretical DCS is a summation over the mechanisms of both KO and
SO. At lower incident electron energy (Fig. 4(a), 10.0 eV above double
ionization threshold), three continuum electrons tend to share the
available energy nearly equivalently. With increasing incident electron
energy, however, one of the three electrons tend to possess most of the
available energy with a large probability, which results in two
slower electrons in the process. The higher energy of the incident electron, the
larger probability of the fast electron. The maximal double DCS
decreases with increasing incident electron energy. Similar
conclusion can be drawn for all other channels. To save space, the
double DCS of N$^+$ and O$^+$ are not given here.

\begin{figure}[t]
\begin{center}
\includegraphics[width=\columnwidth]{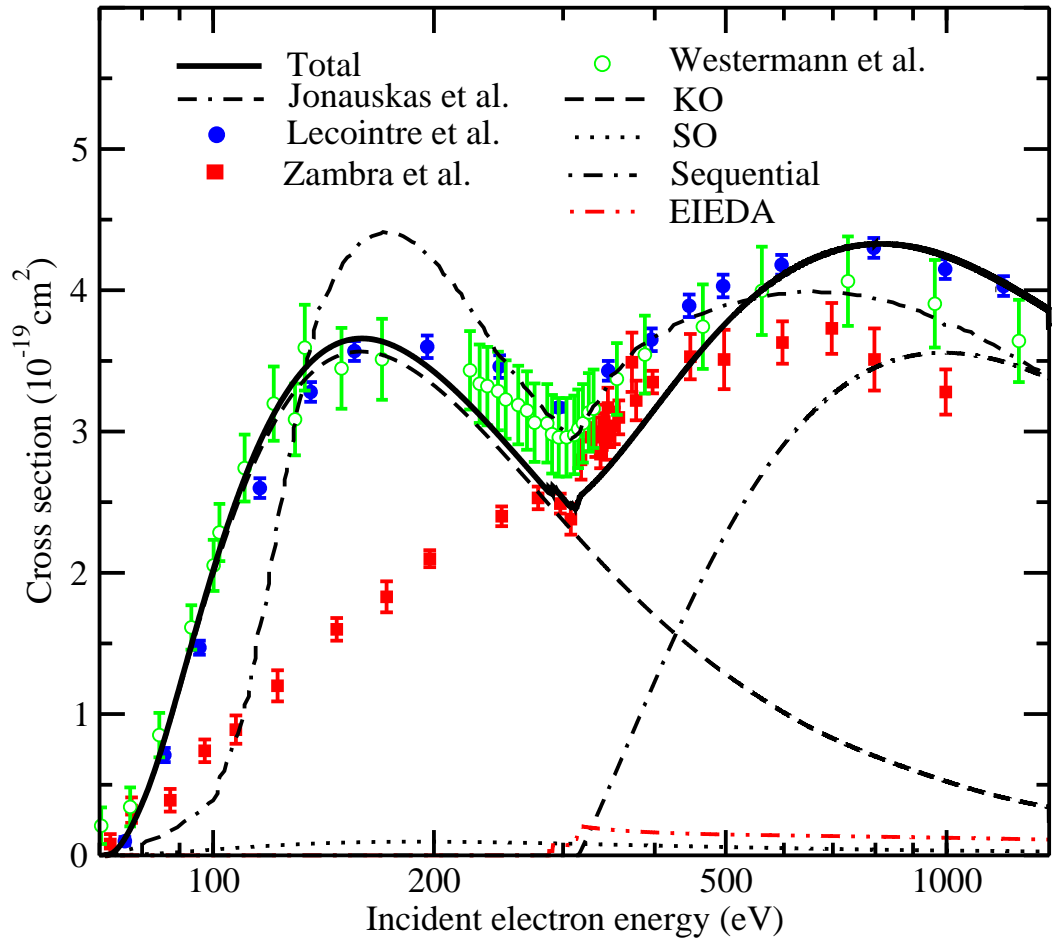}
\caption{Comparison of theoretical electron impact double ionization cross section (the solid line)
of the ground term of C$^+$ with the experimental \cite{Lecointre,Zambra,Westermann}
and theoretical results \cite{Jonauskas2014}. The direct double ionization cross section contributed
by KO and SO mechanisms and indirect processes of sequential and electron impact excitation double autoionization (EIEDA)
are given separately to evaluate their relative importance.}
\end{center}
\end{figure}

In Fig. 5 we compare the electron impact double ionization integral
cross sections for the ground term of C$^+$ with
the experimental \cite{Zambra,Westermann,Lecointre} as well as other
theoretical results \cite{Jonauskas2014}. From the inspection of
this figure, there is an excellent agreement between our theory and
the most recent experimental results obtained by Lecointre {\it et
al.} \cite{Lecointre} from the double ionization threshold to an
incident electron energy of 200 eV. In the intermediate energy range
of 200-500 eV, the experimental cross section is a little larger
than our theoretical prediction. With the further increase of
incident electron energy, we found again a good agreement between
the theoretical and experimental results. Compared with the earlier
measurement of Zambra {\it et al.} \cite{Zambra}, however, a
reasonable agreement is found only at the incident electron energy
range 300-500 eV. In particular, the shape and magnitude near the peak
cross section is completely different. Lecointre {\it et al.}
\cite{Lecointre} measured two peak values of cross section at $\sim$160 and $\sim$900
eV, while Zambra {\it et al.} \cite{Zambra} measured only one peak
value at $\sim$700 eV. The only available theoretical work available in the literature
obtained by Jonauskas {\it et al.} \cite{Jonauskas2014} predicted a larger peak
cross section than our calculation and all available experimental measurements \cite{Zambra,Westermann,Lecointre}.
A large discrepancy can also found near the ionization threshold for the total
double ionization cross section obtained by Jonauskas {\it et al.} \cite{Jonauskas2014}.
\begin{figure}[t]
\begin{center}
\includegraphics[width=\columnwidth]{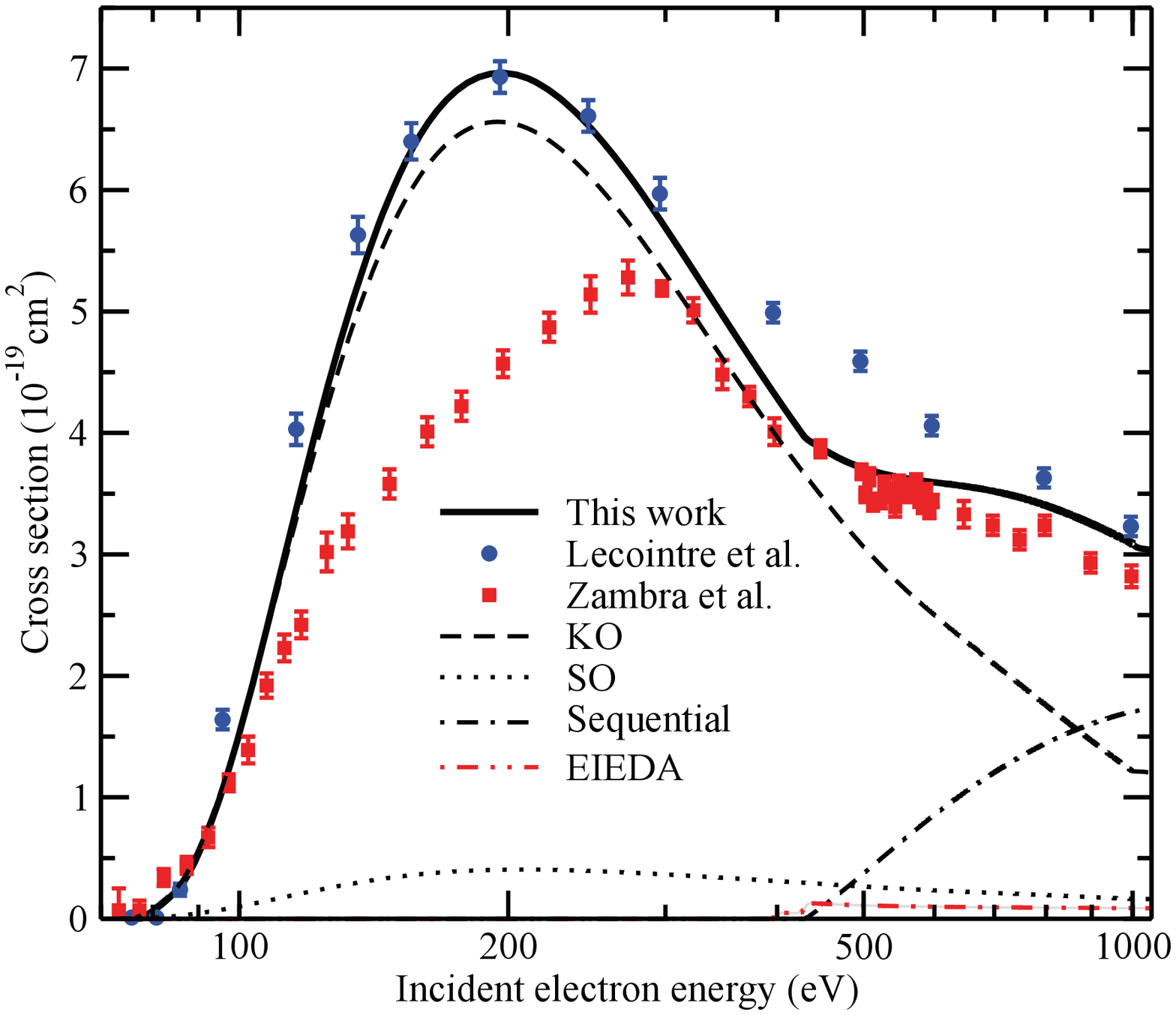}
\caption{As in Fig. 5 but for the electron impact double ionization of the ground term of N$^+$.}
\end{center}
\end{figure}
\begin{figure}[t]
\begin{center}
\includegraphics[width=\columnwidth]{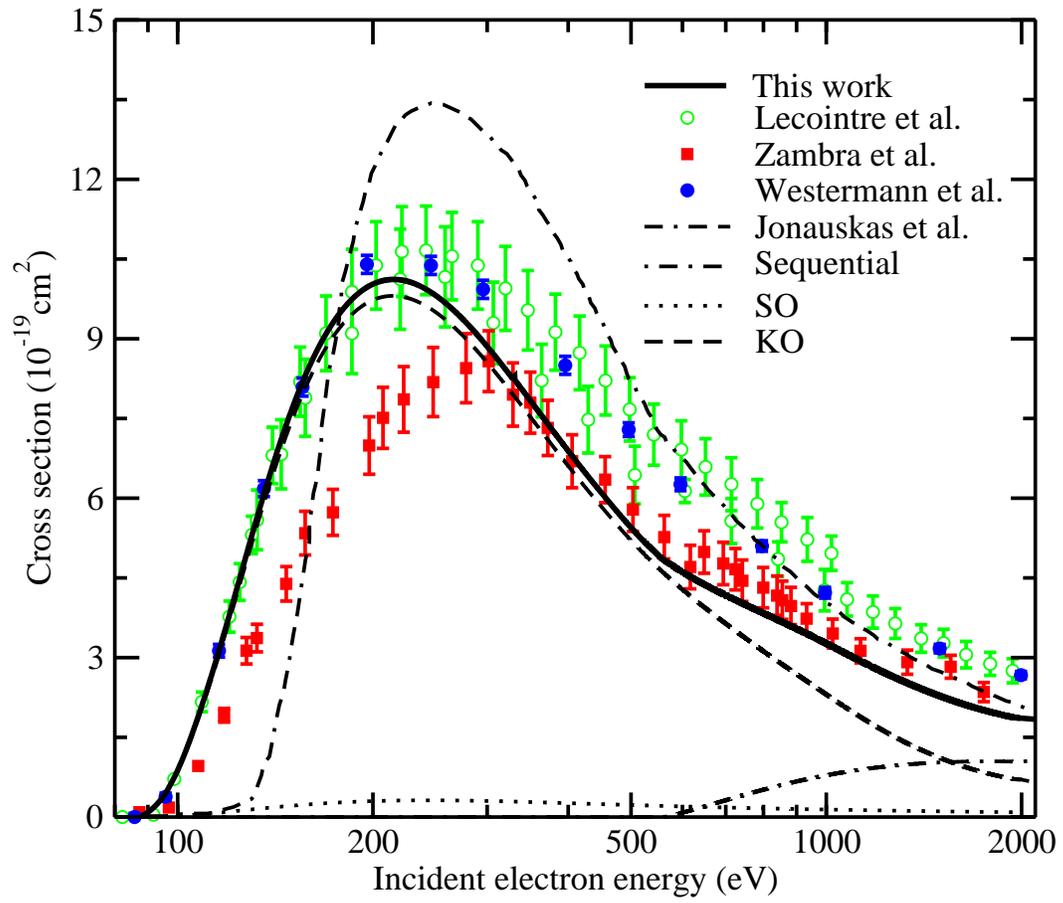}
\caption{As in Fig. 5 but for the electron impact double ionization of the ground term of O$^+$.}
\end{center}
\end{figure}

To have a better understanding of this double ionization process, we give in Fig. 5
the respective contributions of different mechanisms including the
direct and indirect processes. The direct double ionization
refers to contributions of the mechanisms of KO and SO and indirect
processes include sequential double ionization (electron impact
single ionization of $1s$ and subsequently autoionization) and
electron impact excitation of $1s$ and then double autoionization. Obviously, the cross sections are dominated by the KO
mechanisms of direct double ionization from the double ionization
threshold up to $\sim$300 eV, which is perfectly predicted by our
theoretical approach. With increasing incident electron energy up to $\sim$315 eV,
the $1s$ ionization channels are opened and the indirect
ionization processes begin to play a role. Above $\sim$430 eV, the
cross section contributed by the indirect processes becomes larger
than that of direct double ionization channels. On the other hand, the cross
section contributed by the KO mechanism decreases more evidently
than SO with increasing incident electron energy until the latter becomes larger than the former at a very high incident electron energy.

\begin{table*}[tbp]
\caption{\label{energies} Single and double ionization potentials (IP) and 1s single IP (in units of eV) of
the ground terms of C$^{+}$, N$^{+}$, and O$^{+}$ and comparison with the experiment \cite{NIST}. }
\begin{ruledtabular}
\begin{tabular}{llccccc}
\multicolumn{1}{c}{Ions}  & \multicolumn{1}{c}{Ground term} & \multicolumn{2}{c}{Single IP}   & \multicolumn{2}{c}{Double IP}  & \multicolumn{1}{c}{1s IP}\\\cline{3-4} \cline{5-6}
\multicolumn{1}{c}{} & \multicolumn{1}{c}{}  & \multicolumn{1}{c}{This work} & \multicolumn{1}{c}{NIST \cite{NIST}} & \multicolumn{1}{c}{This work} & \multicolumn{1}{c}{NIST \cite{NIST}}  & \multicolumn{1}{c}{} \\\hline
C$^{+}$  & $1s^22s^22p$ $^2P^o$  &24.96 &24.38 &72.36 &72.27 &314.55 \\
N$^{+}$  & $1s^22s^22p^2$ $^3P$  &30.08 &29.60 &76.94 &77.05 &433.07  \\
O$^{+}$  & $1s^22s^22p^3$ $^4S^o$  &35.13 &35.12 &88.30 &90.06 &565.77 \\
\end{tabular}
\end{ruledtabular}
\end{table*}
In Figs. 6 and 7 we compare the theoretical electron impact double ionization integral cross sections
of the ground term of N$^+$ and O$^+$ with the experimental results \cite{Zambra,Lecointre}. The single and double
ionization potentials of these two ions are listed in table 1 along
with the experimental values \cite{NIST}. The contributions to the
cross section from different mechanisms including the direct and
indirect processes are given to evaluate their respective importance. For O$^+$,
theoretical results obtained by Jonauskas {\it et al.}
\cite{Jonauskas2014} are also given. For both ions, there is a general good agreement between
our theoretical cross sections and the most recent experimental
results of Lecointre {\it et al.} \cite{Lecointre}, yet a large
discrepancy is found with the earlier measurement of Zambra {\it et
al.} \cite{Zambra}. Near ionization threshold, the cross sections
are dominated by the KO mechanisms of direct double ionization,
while SO gradually plays a role with increasing incident energy.
Jonauskas {\it et al.} \cite{Jonauskas2014} overestimated the double
ionization cross section around the peak value and their theoretical
double ionization threshold is also larger than the experimental
value. Our results represent the best understanding on the electron
impact double ionization of C$^{+}$, N$^+$ and O$^+$. The
contributions from various direct and indirect double ionization
mechanisms are identified for the first time.

It is interesting to have a comparison of the cross sections of the
three ions of C$^{+}$, N$^+$ and O$^+$. Although these ions differ
by one more electron from each other, there is a large
difference for their integral cross sections. At lower incident
electron energy near the threshold, the peak cross section increases
from 3.9 Mb for C$^+$ to 6.5 Mb for N$^+$ and to 10.8 Mb for O$^+$,
which are dominantly contributed by the KO mechanism of direct
double ionization. However, the cross sections contributed by the
indirect processes decreases fast with the increase of atomic
number from C$^+$ to O$^+$.

\subsection{Double and triple Auger decay}

\begin{figure}[b]
\begin{center}
\includegraphics[width=\columnwidth]{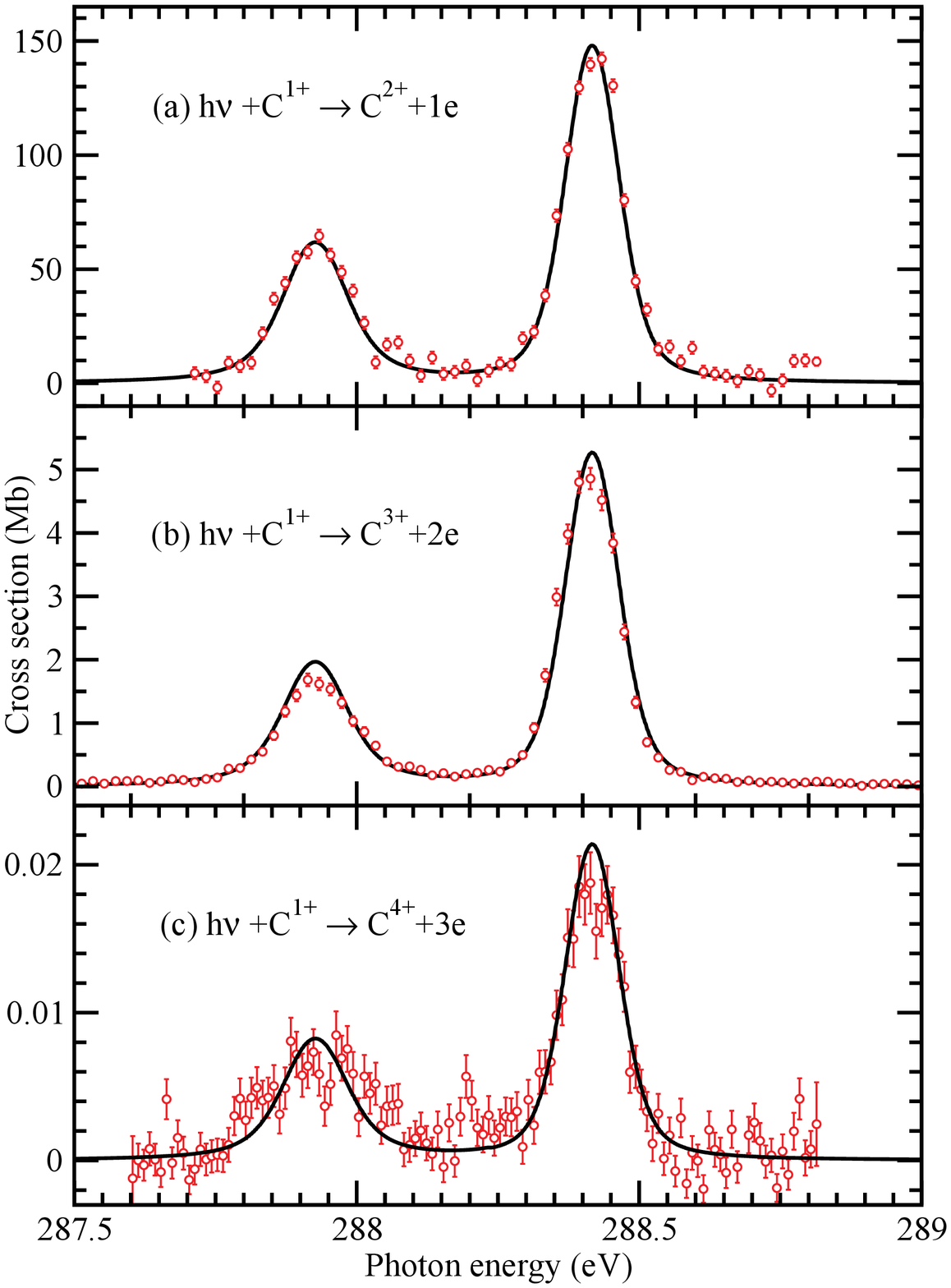}
\caption{Comparison of theoretical (this work, solid line) and experimental
(open circles with error bars) \cite{Muller} cross sections
for the single (a), double (b), and triple (c) ionization of C$^{+}$ ions by a single photon
near the K-shell resonances of $1s2s^22p^2$ $^2D$ and $^2P$. The instrument bandwidth of 92 meV in the
experiment has been included in the theoretical results. To have a better comparison, we have shifted the resonance positions toward
lower photon energy direction by 0.6 eV and 0.7 eV for $^2D$ and $^2P$ terms, respectively.}
\end{center}
\end{figure}

Next we apply our developed theory to investigate the direct double and
triple Auger decay processes. We take the Auger decay of the K-shell excited states
$1s2s^22p^2$ $^2D$ and $^2P$ of C$^{+}$ as examples to illustrate the effectiveness of the approach.
The direct double and triple Auger decay of these autoionization states have been experimentally observed
through measuring the single, double and triple photoionization
cross section by M\"{u}ller {\it et al.} \cite{Muller} and
theoretically investigated by Zhou {\it et al.} \cite{Zhou}. In
table 2 we give the decay rates of the dominant pathways for
the single Auger decay (SAD), DDAD and DTAD of $1s2s^22p^2$ $^2D$
and $^2P$. Through the states of $1s^22s^x2p^y$ ($x+y$=2) these
K-shell excited states decay to $1s^22s^x2p^y$ ($x+y$=1) of C$^{3+}$
in the DDAD. All these paths result in the production of the ground
state $1s^2$ of C$^{4+}$ in the DTAD process. By assuming that the
signatures in the double and triple photoionization of the ground term $1s^22s^22p$ of C$^{+}$ originate solely
from the DDAD and DTAD processes of $1s2s^22p^2$ $^2D$ and $^2P$
terms, the resulting double and triple photoionization cross sections are
shown in Fig. 8. A comparison with the experimental
data \cite{Muller,Muller6} shows that a good agreement is found
between the theory and experiment for the single, double, and triple
photoionization cross sections. Such a good agreement
confirmed that the signatures in the double and triple
photoionization observed in the experiment \cite{Muller} are solely
due to the DDAD and DTAD of the K-shell states. On the other hand,
it also suggests that the interference effects are in general
negligible between SO and KO in the DDAD and between the four
pathways in the DTAD processes.

\begin{figure}[b]
\begin{center}
\includegraphics[width=\columnwidth]{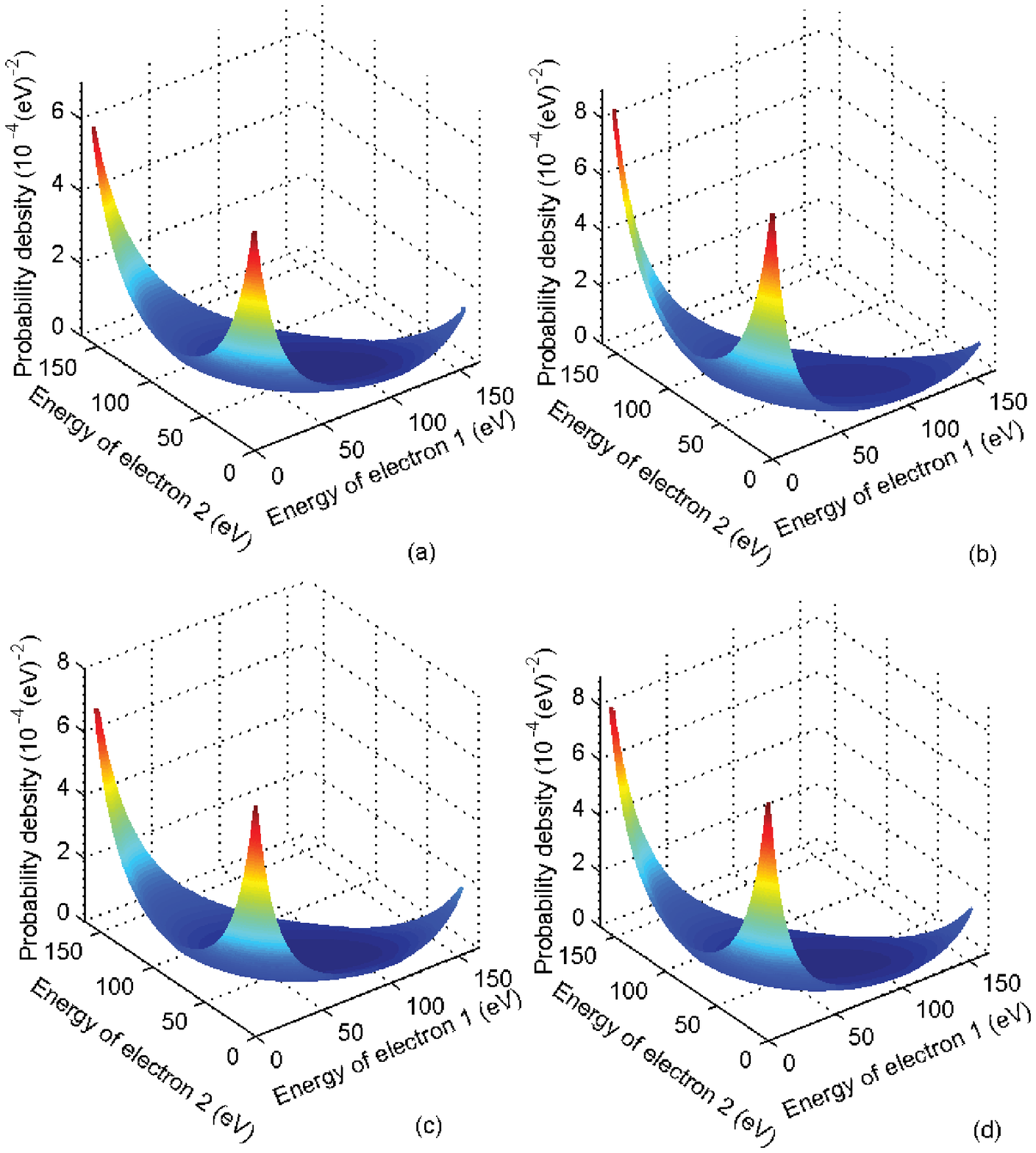}
\caption{Total probability density as a function of energies of two continuum electrons in the DTAD
processes of C$^{+}$ $1s2s^22p^2$ $^2D$ term for the four strongest pathways:
(a) C$^{2+}$ $1s^22s^2$ $^1S$ $\rightarrow$ C$^{3+}$ $1s^22s$ $^2S$ $\rightarrow$ C$^{4+}$ $1s^2$,
(b) C$^{2+}$ $1s^22s2p$ $^1P^o$ $\rightarrow$ C$^{3+}$ $1s^22s$ $^2S$ $\rightarrow$ C$^{4+}$ $1s^2$,
(c) C$^{2+}$ $1s^22s2p$ $^1P^o$ $\rightarrow$ C$^{3+}$ $1s^22p$ $^2P^o$ $\rightarrow$ C$^{4+}$ $1s^2$,
and (d) C$^{2+}$ $1s^22s^02p^2$ $^1D$ $\rightarrow$ C$^{3+}$ $1s^22p$ $^2P^o$ $\rightarrow$ C$^{4+}$ $1s^2$.}
\end{center}
\end{figure}

\begin{table*}
\caption{Auger decay rates ($s^{-1}$) of the dominant pathways for the SAD (A$^1$), DDAD (A$^2$), and DTAD (A$^3$) processes from the initial terms of C$^{+}$ $1s2s^22p^2$ $^2D$ and $^2P$. The intermediate terms of C$^{2+}$ and C$^{3+}$ are given to specify the pathways. Figures in brackets indicate powers of ten. }
\begin{ruledtabular}
\begin{tabular}{llclccc}
\multicolumn{1}{l}{Initial C$^{+}$}  & \multicolumn{1}{c}{C$^{2+}$} & \multicolumn{1}{c}{A$^1$} & \multicolumn{1}{c}{C$^{3+}$}  & \multicolumn{1}{c}{A$^2$}  & \multicolumn{1}{c}{Final C$^{4+}$}  & \multicolumn{1}{c}{A$^3$}  \\\hline
$1s2s^22p^2$ $^2D$ & $1s^22s^2$ $^1S$ &5.026[13]  &$1s^22s$ $^2S$ &1.388[12] &$1s^2$ &4.086[9] \\
       & $1s^22s2p$ $^3P^o$    &7.641[12]  &$1s^22s$ $^2S$        &1.620[11] &$1s^2$ &5.288[8]\\
       & $1s^22s2p$ $^3P^o$    &           &$1s^22p$ $^2P^o$      &1.099[11] &$1s^2$ &4.872[8]\\
       & $1s^22s2p$ $^1P^o$    &3.546[13]  &$1s^22s$ $^2S$        &7.360[11] &$1s^2$ &2.412[9]\\
       &$1s^22s2p$ $^1P^o$     &           &$1s^22p$ $^2P^o$      &5.528[11] &$1s^2$ &2.385[9] \\
       &$1s^22s^02p^2$ $^1D$ &5.526[13]  &$1s^22s$ $^2S$          &1.903[11] &$1s^2$ &6.365[8] \\
       &$1s^22s^02p^2$ $^1D$ &           &$1s^22p$ $^2P^o$        &2.288[12] &$1s^2$ &9.405[9] \\
       &total                &1.557[14]  &total                   &5.762[12] &total  &2.168[10]\\
$1s2s^22p^2$ $^2P$ & $1s^22s2p$ $^1P^o$ &1.216[13]  &$1s^22s$ $^2S$ &2.530[11] & $1s^2$  &8.534[8]\\
       & $1s^22s2p$ $^1P^o$    &           &$1s^22p$ $^2P^o$        &1.896[11] & $1s^2$  &8.088[8]\\
       & $1s^22s^02p^2$ $^3P$           &5.606[13]  &$1s^22p$ $^2P^o$ &2.399[12] & $1s^2$ &9.993[9]\\
       &total                           &7.670[13]  &total            &3.102[12] & total  &1.194[10]\\
\end{tabular}
\end{ruledtabular}
\end{table*}

\begin{figure}[b]
\begin{center}
\includegraphics[width=\columnwidth]{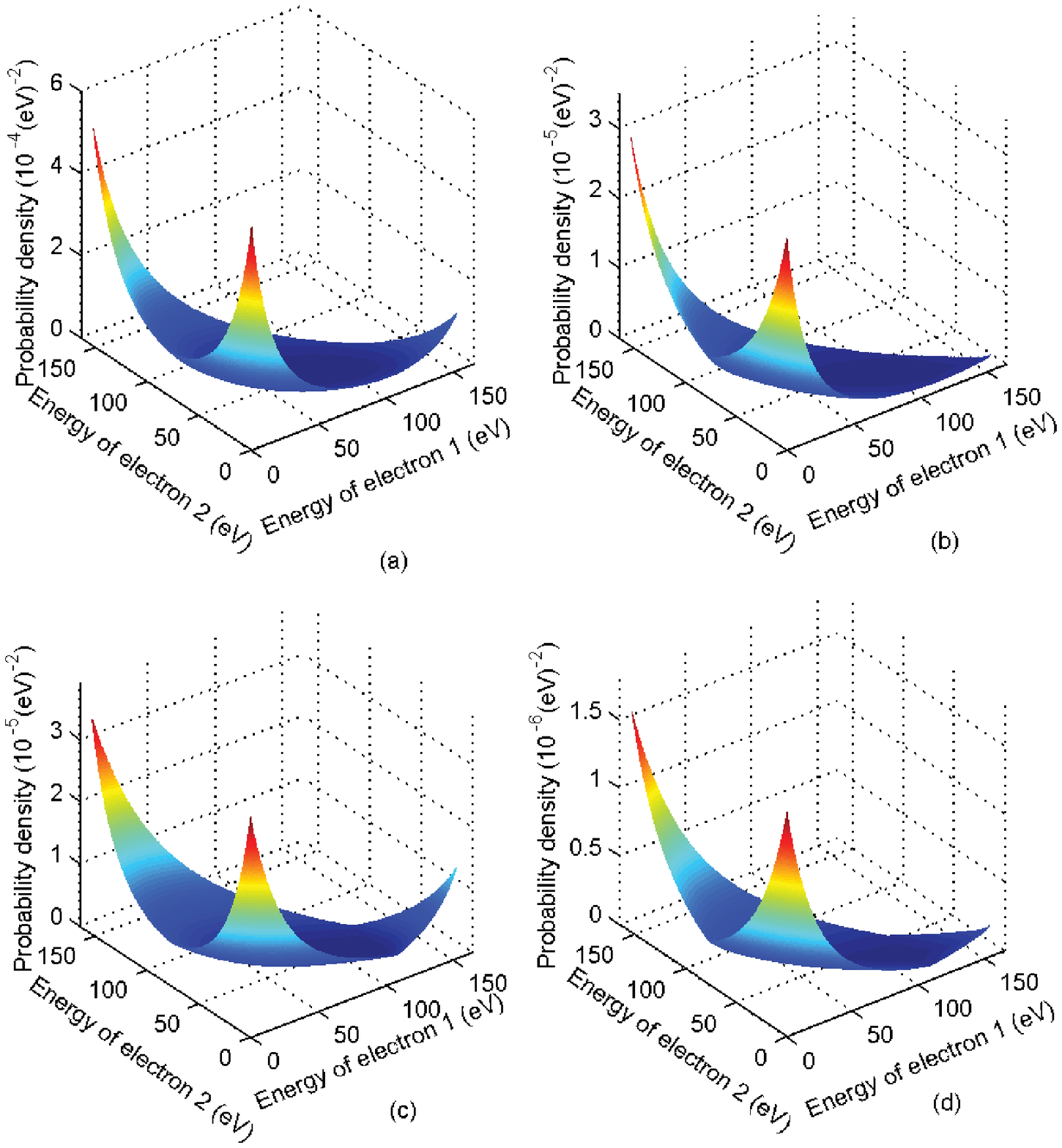}
\caption{Probability density contributed by (a) double KO, (b) KO+SO, (c) SO+KO, and (d) double SO
for the pathway of C$^{+}$ $1s2s^22p^2$ $^2D$ $\rightarrow$
C$^{2+}$ $1s^22s^2$ $^1S$ $\rightarrow$ C$^{3+}$ $1s^22s$ $^2S$ $\rightarrow$ C$^{4+}$ $1s^2$.}
\end{center}
\end{figure}

To gain more insight into the DTAD process, we have also investigated the
energy correlation among the three Auger electrons. Fig. 9 shows the
total probability density over a summation of the four mechanisms as
a function of energies of two Auger electrons for the four strongest
pathways of $1s2s^22p^2$ $^2D$ (the first, fourth, fifth, and
seventh lines in table 2). It is concluded that the DTAD process
favors one faster and two slower electrons. The probability density
is very small for equal sharing of the energy among the three
electrons at a higher photon energy. Fig. 10 shows the probability density for the pathway of
C$^{+}$ $1s2s^22p^2$ $^2D$ $\rightarrow$ C$^{2+}$ $1s^22s^2$ $^1S$ +
e $\rightarrow$ C$^{3+}$ $1s^22s$ $^2S$ + 2e $\rightarrow$ C$^{4+}$
$1s^2$ $^1S$ + 3e (the first line in table 2) contributed by (a)
double KO, (b) KO+SO, (c) SO+KO, and (d) double SO. Among the four
mechanisms, double KO dominates the decay processes.
The probability density of KO+SO and SO+KO is smaller by
more than one order of magnitude and that of double SO is
smaller by more than two orders of magnitude than that of double KO.
\subsection{Double and triple photoionization}

Most researches on DPI are carried out on atomic helium and
helium-like ions (for examples, see \cite{Schneider,Samson,Pattard,Hino} and references therein), which are the simplest
systems for such a process to occur. Atomic lithium is a system with only one more electron
than helium, yet its theoretical description of DPI is much more complicated \cite{Hugo,Colgan,Kheifets,Kheifets1,Colgan1,Hugo,Wehlitz4}.
Various theoretical methods have been developed including convergent close-coupling (CCC) \cite{Kheifets1},
time-dependent close coupling (TDCC) \cite{Colgan1,Colgan}, R-matrix
with pseudostates (RMPSs) \cite{Colgan1} and other theoretical method \cite{Hugo}.
Experimentally, Huang {\it et al.} \cite{Huang} and Wehlitz {\it et al.} \cite{Wehitz1}
measured the double-to-single ratios of photoionization cross sections of atomic Li.
Lithium atom is also the prototype object to investigate the triple photoionization (TPI),
both theoretically \cite{Hugo,Santos,Mikhailov,Wehitz3,Juranic,Emmanouilidou,Pattard1} and experimentally \cite{Wehitz2}.
In order to investigate both the DPI and TPI processes, we chose atomic lithium as
our study object.

\begin{figure}[htbp]
\begin{center}
\includegraphics[width=\columnwidth]{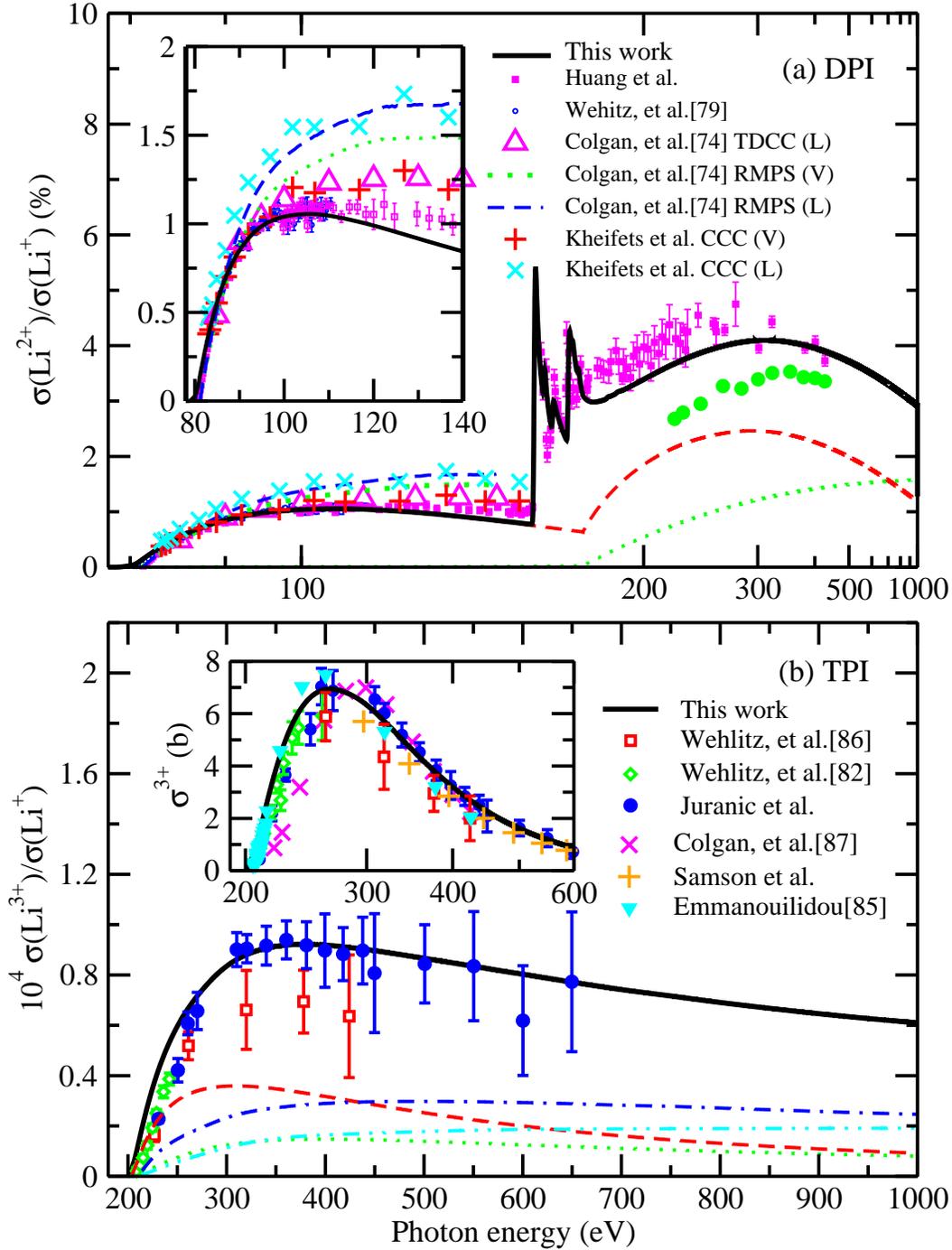}
\caption{(a) Double-to-single ratio of photoionization cross section for atomic lithium compared with available
experimental \cite{Wehitz1,Huang} and theoretical results of the length and velocity gauges of CCC calculations
by Kheifets {\it et al.} \cite{Kheifets1} and the length gauge of TDCC, length and velocity gauges of RMPS
by Colgan {\it et al.} \cite{Colgan1} available in the literature. The calculations of CCC(L), CCC(V), TDCC (L)
, RMPS(L), and RMPS(L) are represented by $\times$, +, magenta $\bigtriangleup$, a blue dashed line, and a green dotted line,
respectively. The dashed and dotted lines represent the cross section contributed by the KO and SO mechanisms.
The inset enlarged the double-to-single ratio from the ionization threshold to a photon energy of 140 eV to have a better comparison
near the double ionization threshold.
(b) Comparison of our theoretical triple-to-single ratio of photoionization cross section of lithium
with available experimental measurements by Wehitz {\it et al.} \cite{Wehitz2,Wehitz3} and
Jurani\'{c} {\it et al.} \cite{Juranic} and theoretical results of Colgan {\it et al.} (magenta $\times$) \cite{Colgan},
Emmanouilidou and Rost (cyan $\bigtriangledown$) \cite{Emmanouilidou}, and Samson {\it et al.} \cite{Samson} (yellow +).
The cross section contributed by the KO+KO, KO+SO, SO+KO, and SO+SO mechanisms
are represented by a red dashed, blue dotted-dashed, dotted-dotted-dashed, and green dotted line,
respectively. In the inset, a comparison is made for the triple photoionization cross section with
available experimental and theoretical results.}
\end{center}
\end{figure}

\begin{figure}[htbp]
\begin{center}
\includegraphics[width=\columnwidth]{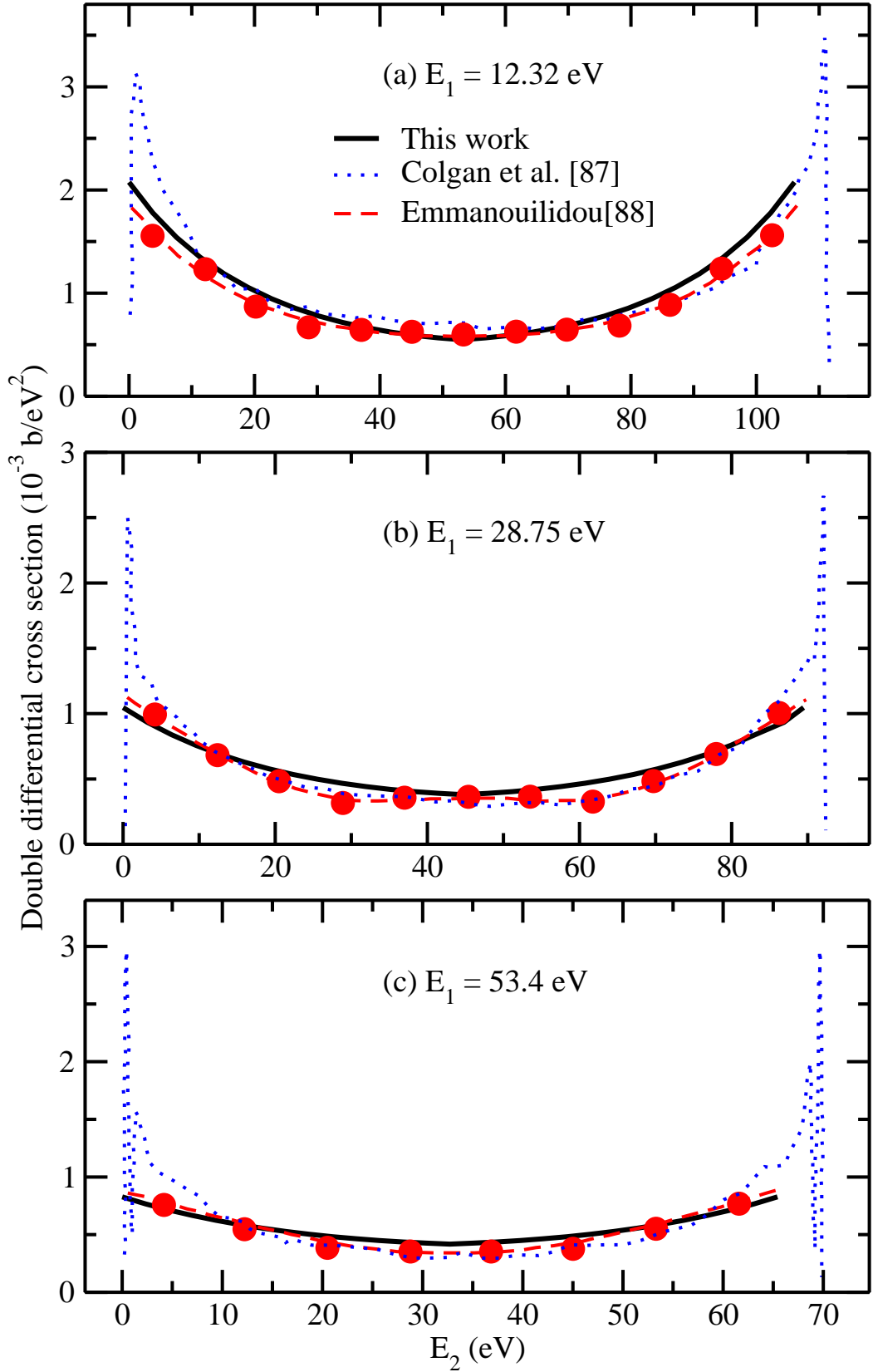}
\caption{Energy resolved double DCS for triple photoionization of the ground state of atomic Li
at an excess energy of 115 eV. By fixing the energy of one electron to be (a) 12.32 eV, (b) 28.75 eV, (c) 53.4 eV,
we show the double DCS as a function of the energy of another continuum electron. Red circles and red
lines represent the theoretical calculations in a quasiclassical framework \cite{Emmanouilidou07} and blue
dotted lines are the theoretical results from Ref. \cite{Colgan06} at a photon energy of 320 eV.}
\end{center}
\end{figure}

The calculated double-to-single ratio of photoionization cross section of
lithium is shown in Fig. 11(a) in a black solid line. The separate contributions from
the mechanisms of KO and SO are given in a dashed and dotted line, respectively.
It can be seen that only direct ionization processes contributed
to the DPI cross section from the ionization threshold to a photon energy of 150.3 eV and
it originates predominantly from the KO mechanism. With the increase of photon energy up to 170.1 eV
(which corresponds to the opening of channel $h\nu+1s^22s \rightarrow 2s^2+e \rightarrow 2s+2e$),
the cross section contributed by SO processes gradually increases and plays a role. At the photon energy of
710.0 eV, the SO and KO mechanisms have an equivalent contribution. With the further increase
of photon energy from 710.0 eV, the SO mechanism is becoming more important than KO and
it dominates in the direct DPI at a higher photon energy. Obviously, there are additional contribution
from the photon energy of 150.3 eV, which can be seen from the rapid increase of the cross section.
We have identified that the additional contribution originates from the indirect DPI processes.
It is caused by the ionization
of $1s$ electron plus excitation of another $1s$, resulting in the production of $1s^0nln'l'$ excited states.
These highly excited states lie above the $1s$ threshold of Li$^{+}$ and they decay by Auger processes
to the doubly ionized states of Li$^{2+}$. The most important indirect pathways originate from $1s^02s^2$,
$1s^02s2p$ and $1s^02s3s$. The first steep increase of the double-to-single ratio is due to $1s^02s^2$
with a threshold energy of 150.3 eV and to $1s^02s2p$ at a photon energy of 150.5 eV. The second increase located at 163.2 eV
originates from the opening of channel $1s^02s3s$.

We compare our theoretical double-to-single ratio of photoionization cross section with the available
experimental \cite{Wehitz1,Huang} and theoretical results \cite{Kheifets1,Colgan1} in the literature.
From the inspection of Fig. 11(a), we found an excellent agreement between our theoretical results and both experimental
measurements carried out by Wehitz {\it et al.} \cite{Wehitz1} and Huang {\it et al.} \cite{Huang} in photon energy range
from the double ionization threshold to 120 eV. From the photon energy of 120 eV to the opening of indirect channel
of $1s^02s^2$, there is also a good agreement with the experiment of Huang {\it et al.} \cite{Huang}.
To have a better comparison, the double-to-single ratio from the ionization threshold to 140 eV
was enlarged in the inset along with available theoretical results obtained by Kheifets {\it et al.} \cite{Kheifets1}
(length and velocity gauges of CCC calculations) and Colgan {\it et al.} \cite{Colgan1} (length gauge of
TDCC, length and velocity gauges of RMPS). Obviously, our theoretical results are closer to the experimental
values of Wehitz {\it et al.} \cite{Wehitz1} and Huang {\it et al.} \cite{Huang} than other theories.
At a higher photon energy, our theory agrees reasonably well with the only available experiment of
Wehitz {\it et al.} \cite{Wehitz1}. The theoretical results obtained by Colgan {\it et al.} \cite{Colgan1}
using a velocity gauge of RMPS underestimated the double-to-single ratio. The underestimation is due to that
they did not include the contribution of indirect double ionization processes in their calculations.

The ratio of triple-to-single photoionization cross section is presented
in Fig. 11(b) in a solid line. The contributions of KO+KO, KO+SO, SO+KO, and SO+SO mechanisms
are represented by a red dashed, blue dotted-dashed, dotted-dotted-dashed, and green dotted line,
respectively. At a lower photon energy near the ionization threshold, the largest
cross section is contributed by the KO+KO mechanism. With increasing photon energy,
the contributions of KO+SO, SO+KO, and SO+SO mechanisms become more and more important.
Comparison with the experimental results obtained by Wehitz {\it et al.} \cite{Wehitz2,Wehitz3}
and Jurani\'{c} {\it et al.} \cite{Juranic} shows that a good agreement is found between our theoretical
calculations and the latest experimental measurements of Jurani\'{c} {\it et al.} \cite{Juranic}.
In the inset of Fig. 11(b), we compare our theoretical triple photoionization cross section of the ground state of
atomic Li with available theoretical results reported in the literature obtained by Colgan {\it et al.} \cite{Colgan},
Emmanouilidou and Rost \cite{Emmanouilidou}, and Samson {\it et al.} \cite{Samson} and
experimental measurements by Wehitz {\it et al.} \cite{Wehitz2, Wehitz3} and Jurani\'{c} {\it et al.} \cite{Juranic}
in the photon energy range from triple ionization threshold to 600 eV.
Our results are in a good agreement with the latest experimental results \cite{Juranic}
and theoretical calculations of Emmanouilidou and Rost \cite{Emmanouilidou}.
Reasonable agreements are found with other experimental and theoretical results.

In Fig. 12 we compare our calculated energy resolved double DCS
with other available theoretical results \cite{Colgan06,Emmanouilidou07} for an excess energy (sum of energy
of the three ejected electrons) of 115 eV after triple photoionization. The double DCS is
given as a function of energy of another outgoing electron by fixing the energy of one electron
to be 12.32eV, 28.75 eV, and 53.4 eV, respectively. Colgan {\it et al.} \cite{Colgan06} calculated energy
resolved DCS of triple photoionization of Li using non-perturbative TDCC method.
Emmanouilidou \cite{Emmanouilidou07} computed double energy resolved DCS in a quasiclassical framework.
Note that there is a little difference for the energy of the second electron in the work of Colgan {\it et al.} \cite{Colgan06}
for each case. In the calculation of Colgan {\it et al.} they chose the photon energy to be 320 eV
which corresponds to an excess energy of $\sim$120.0 eV. The energy resolved double DCS shows a "U-shape" in all calculations.
The larger the excess energy,
the more pronounced the unequal energy sharing between the three electrons. A reasonable agreement is found
between three theoretical calculations except for Colgan {\it et al.} \cite{Colgan06} at the lowest and the highest electron energies.

In what follows we would like to say a few words on the interference effects in
the multielectron processes. From the detailed investigations on the typical
multielectron processes shown in the above three subsections we
conclude that the coherence characteristics is mostly lost in the
energy resolved DCS and integral cross section. However, the momenta
of the ejected electrons continue to be correlated and could be measured with a very high
resolution. If the quantum interference can only be found with
a resolution beyond the experimental capability, the
quantum features of the processes exist only in the calculation of
the cross sections or the reaction rates. The partial loss of
coherence of particles due to the environment coupling by Coulomb
interaction has experimentally explored by Akoury {\it et al.}
\cite{sci2007} in the DPI of molecular hydrogen. We suggest that
the loss of phases of matter wave for the later ionized electrons (not
the first-ionized electron) is a general phenomenon, not only for
the multielectron processes discussed in this work but also for
those such as in strong field ionization \cite{Liuxj}. Liu {\it et
al.} \cite{Liuxj} stated that the first-ionized electron in
laser-induced non-sequential multiple ionization either leaves
immediately after re-collision or joins the other electrons to form
a thermalized complex after a time delay of a few hundred
attoseconds. Our study will shed light on the quantum manipulation
of many particle processes and systems such as the entangled
many-particle states in the quantum computing.
\section{Conclusion}
In summary, we proposed an accurate and practical theoretical formalism on
the correlated dynamical multielectron processes by fully taking
into account of the advantage of the coherence characteristics of
the bound and continuum electrons in the initial and final states.
Detailed expressions are obtained to investigate the energy
correlations among the continuum electrons and energy resolved
differential and integral cross sections in the electron impact
direct double ionization processes according to the separation of KO
and SO mechanisms. Extension to the higher order multielectron processes
is straight forward by adding the contributions of the following KO
and SO processes. The method is firstly applied to the electron impact
double ionization processes of the ground terms of C$^+$, N$^+$, and O$^+$.
The contributions of different mechanisms including both direct and indirect processes are identified for the first
time. Then we employ the method to investigate the direct double and triple
Auger decay in the K-shell excited states of C$^{+}$ $1s2s^22p^2$
$^2D$ and $^2P$. We nicely explain the recent experimental
observation of direct triple Auger decay for the first time and confirmed that the signatures of
the resonances found in the triple photoionization of C$^{+}$ originates
solely from the DTAD process of the K-shell resonant states. The calculated energy correlation
among the three continuum electrons in both processes of electron
impact double ionization and direct triple Auger decay support a
physical scenario of a faster and two slower electrons. At last, the
approach is applied to investigate the double and triple
photoionization of atomic lithium and comparisons are made with
available experimental and theoretical results. Our results show that the proposed
approach is accurate and effective in the treatment of atomic
multielectron processes. It is natural to extend the present
theoretical formalism to the higher more-electron processes. It should also be tractable to extend
the present approach to more complex systems such as molecules and clusters.

\begin{acknowledgments}

This work was supported by the National Natural
Science Foundation of China under Grant Nos. 11674394 and 11674395.

\end{acknowledgments}

\end{document}